\documentclass[aps,prl,twocolumn,groupedaddress,floatfix]{revtex4-1}
\usepackage{graphicx,amsmath,amssymb}
\usepackage[dvips]{color}

\newcommand\be{\mathbf{e}}
\newcommand\bk{\mathbf{k}}
\newcommand\bx{\mathbf{x}}
\newcommand\bu{\mathbf{u}}

\newcommand\kgpermc{\nobreak\mbox{$\;$kg\,m$^{-3}$}}

\newcommand\mNperm{\nobreak\mbox{$\;$mN\,m$^{-1}$}}
\newcommand\mperssq{\nobreak\mbox{$\;$m\,s$^{-2}$}}
\newcommand\msqpers{\nobreak\mbox{$\;$m$^2$\,s$^{-1}$}}
\newcommand\Hz{\nobreak\mbox{$\;$Hz}}
\newcommand\mm{\nobreak\mbox{$\;$mm}}

\newcommand\s{\nobreak\mbox{$\;$s}}

\begin{document}

\title{Alternating hexagonal and striped patterns in Faraday 
surface waves}

\author{Nicolas P\'erinet}
\affiliation{Faculty of Science,
University of Ontario Institute of Technology (UOIT),
Oshawa, Ontario, Canada, L1H 7K4}
\author{Damir Juric}
\affiliation{LIMSI-CNRS (UPR 3251), B.P. 133, 91403 Orsay France}
\author{Laurette S. Tuckerman}
\affiliation{
PMMH (UMR 7636 CNRS - ESPCI - UPMC Paris 6 - UPD Paris 7), 
10 rue Vauquelin, 75005 Paris France}
\email{laurette@pmmh.espci.fr}

\begin{abstract}
A direct numerical simulation of Faraday waves is carried out in 
a minimal hexagonal domain. 
Over long times, we observe the alternation of patterns we call 
quasi-hexagons and beaded stripes.
The symmetries and spatial Fourier spectra of these patterns are analyzed. 
\end{abstract}

\maketitle

\begin{figure}
\centerline{\includegraphics[width=9cm,clip]{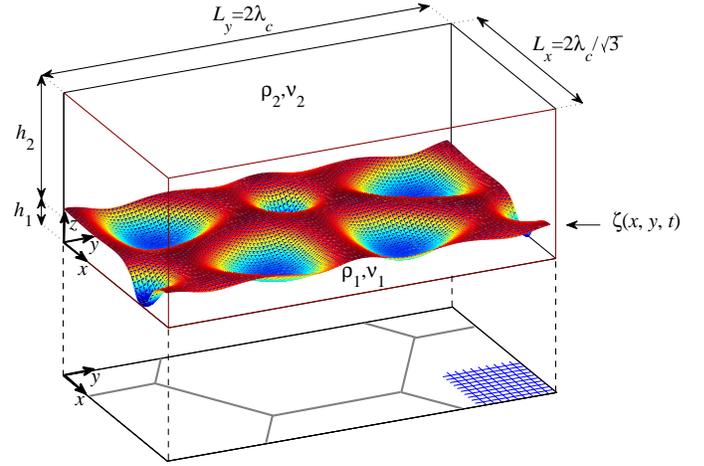}}
\caption{Above: instantaneous realization of hexagaonal Faraday waves in a
  domain of size $L_x\times L_y \times (h_1+h_2)$.  The density and viscosity
  are $\rho_2$, $\nu_2$ above the free surface at $z=\zeta(x,y,t)$; below,
  they are $\rho_1$ and $\nu_1$.  The displacement amplitude of $\zeta$ is
  comparable to $h_1$; at its lowest points, $\zeta$ almost touches the lower
  boundary.  Below: projection of the horizontal doubly periodic rectangular
  domain of dimensions $L_x\times L_y=2\lambda_c/\sqrt{3} \times 2\lambda_c$
  shows that it is compatible with a hexagonal lattice.  The minima of $\zeta$
  (above) are located at the vertices and centers of the hexagons (below). The
  right corner represents a portion of the rectangular computational grid; the
  actual grid is twice as fine in each direction.}
\label{fig:domaine_complet}
\end{figure}

The Faraday instability \cite{Far1831} describes the generation of surface
waves between two superposed fluid layers subjected to periodic vertical
vibration. Although these waves usually form crystalline patterns,
i.e.~stripes, squares, or hexagons, they can form more complicated
structures such as quasicrystals or superlattices \cite{CAL1992,EF1993,KPG1998,RSS2012}. 
We have
recently carried out the first three-dimensional nonlinear simulations of the
Faraday instability \cite{Perinet_jfm_2009} and have reproduced the square and
hexagonal patterns seen in \cite{KEMW2005,*KEMW2009} with the same physical
parameters; see Fig.~\ref{fig:domaine_complet}.  However, the hexagonal
pattern is not sustained indefinitely, but is succeeded by recurrent
alternation between quasi-hexagonal and beaded striped patterns.  This
long-time behavior is the subject of the present study.

\begin{figure*}
\centerline{\includegraphics[width=\textwidth]{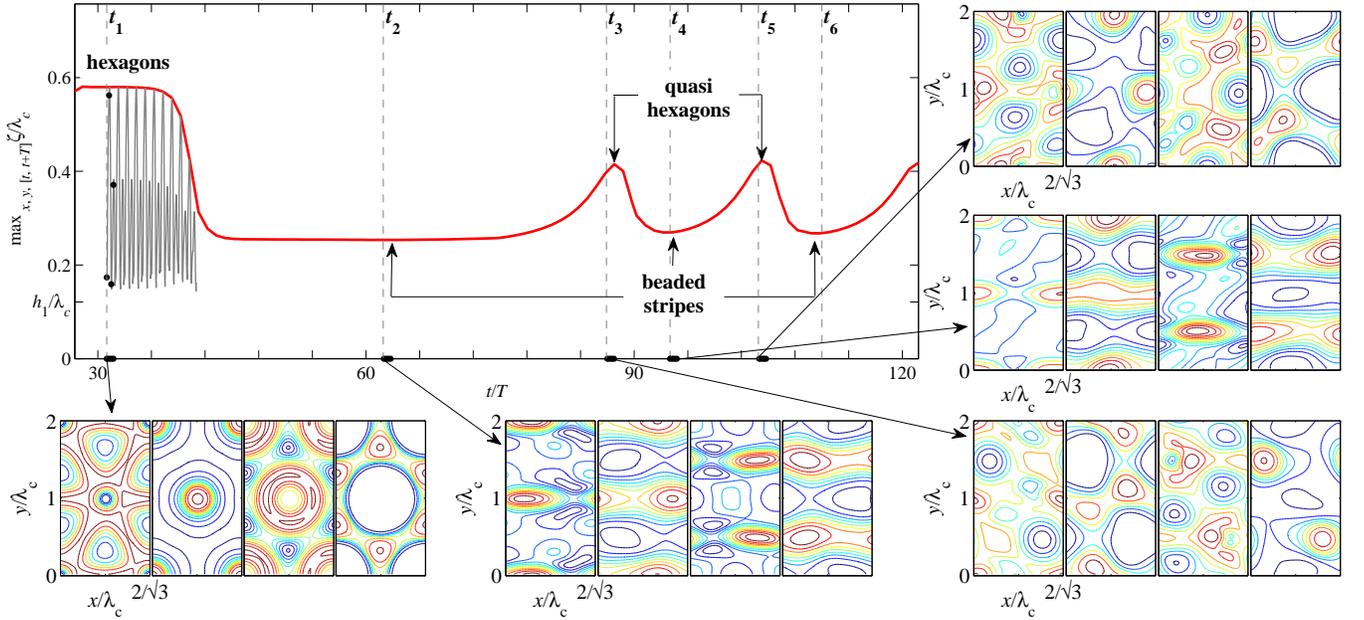}}
\vspace*{-0.2cm}
\caption{Maximum interface heights $\max_{x,y}\zeta(x,y,t)$ (rapidly
  oscillating curve) and $\max_{x,y,[t,t+T]}\zeta(x,y,t)$ (smooth envelope).
  Surrounding visualizations show instantaneous contour plots of
  $\zeta(x,y,t)$ at times indicated by black dots on rapidly oscillating curve
  and along abscissa. 
Red/blue colors regions of maximum/minimum interface height.
The color map of each contour plot is scaled independently of the
  others.  Over the large white areas, the interface is very close to the
  bottom and almost flat.  Visualizations shown at $t_i+jT/4$ for $j=0,1,2,3$
  (i.e.~over one subharmonic period) for hexagons at $t_1$, symmetric and
  nonsymmetric beaded stripes at $t_2$ and $t_4$ and quasi-hexagons at $t_3$
  and $t_5$.  
}\label{fig:deltazeta_complet}
\end{figure*}

The first detailed spatio-temporal experimental measurements of the interface
height $z=\zeta(x,y,t)$ were undertaken by \cite{KEMW2005,*KEMW2009}.  Their
optical technique required the two fluid layers to have the same refractive
index, which led them to use fluids of similar viscosities and densities:
$\rho_1 =1346\kgpermc$, $\nu_1=5.35\times 10^{-6} \msqpers$, $\rho_2
=949\kgpermc$, $\nu_2=2.11\times 10^{-5}\msqpers$ and surface tension
$\sigma=35\mNperm$.  These parameters, especially the density ratio
$\rho_2/\rho_1=0.7$, differ markedly from most studies of Faraday waves, which
use air above either water or silicone oil and so have $\rho_2/\rho_1\approx
0.001$.  At rest, the heavy and light fluids occupy heights of $h_1=1.6\mm$
and $h_2=8.4\mm$, respectively.  The imposed vibration has frequency $f=12\Hz$
and the Faraday instability leads to subharmonic standing waves, so that
$\zeta(x,y,t)$ oscillates with period $T=2/f=0.1666\s$.  Floquet analysis
\cite{KT1994} for these parameters yields a critical wavelength of $\lambda_c
=2\pi/k_c=13.2\mm$, with which the experiments show close agreement
\cite{KEMW2005,*KEMW2009}. 
Thus, another atypical feature of this parameter regime
is that $h_1 \ll\lambda_c$; see Fig.~\ref{fig:domaine_complet}.  Floquet
analysis also yields a critical acceleration of $a_c=25.8 \mperssq=2.63\:
g$. Our simulations are carried out at $a=38.0\mperssq =3.875\:g =
1.473\;a_c$, for which hexagons were observed experimentally
\cite{KEMW2005,*KEMW2009}. 

We summarize our formulation and the numerical methods used to compute the
fluid motion; see \cite{Perinet_jfm_2009} for a more detailed description.
Our computations use a single-fluid model, representing the velocity $\bu$ and
pressure $p$ over the whole domain on a staggered MAC mesh \cite{HW1965} which
is fixed and uniform.  The viscosity and density are variable, taking the
values $\nu_1$, $\rho_1$ for the denser lower fluid, $\nu_2$, $\rho_2$
for the lighter upper fluid and varying over a few gridpoints at the
interface. The moving interface, defined by $z=\zeta(x,y,t)$, is computed by a
front-tracking~\cite{T&al2001}/immersed-boundary~\cite{Pes1977} method on a
semi-Lagrangian triangular mesh which is fixed in the horizontal $x$ and $y$
directions and moves along the vertical direction $z$. The interface is
advected and the density and viscosity fields updated. The capillary force is
computed locally on the Lagrangian mesh and included in the Navier-Stokes
equations, which are solved by a projection method.  The computations are
carried out in the oscillating reference frame of the container by adding a
time-periodic vertical acceleration $a \sin(2\pi f\: t)\be_z$ to the equations
of motion.  No-slip boundary conditions are imposed at the top and bottom
boundaries, while periodic boundary conditions are used at the vertical
boundaries. 

The horizontal dimensions of the domain are chosen to accomodate a hexagonal
pattern.  We take $L_x=2\lambda_c/\sqrt{3}$ and $L_y=2\lambda_c$, as shown in
Fig.~\ref{fig:domaine_complet}, so that large-scale
spatial variations are inaccessible.  This domain is also compatible
with striped or rectangular patterns, as will be discussed below.
The simulations were run with a spatial resolution of $N_x\times N_y \times
N_z = 58 \times 100 \times 180$. Each horizontal rectangle is subdivided into
64 triangles to represent the interface.  To validate the spatial
discretization, we repeated the simulations with a finer resolution of
$N_x\times N_y \times N_z = 75 \times 125 \times 225$.  Although small
quantitative changes were seen, the dynamics remained qualitatively unchanged.
The timestep is limited by the advective step, taking values varying between
$T/24\,000$ for a hexagonal pattern and $T/4000$ for a beaded striped
pattern. This makes the simulation of behavior over many subharmonic periods
extremely time-consuming.

\begin{figure*}
\includegraphics[width=17cm]{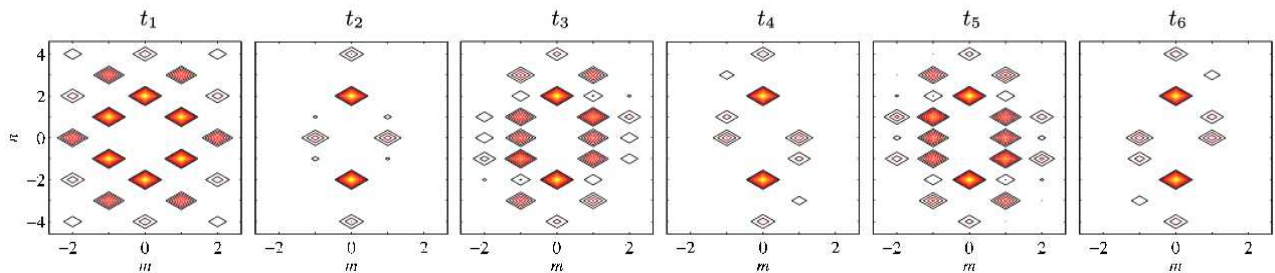}
\vspace*{-0.3cm}
\caption{Time-filtered spatial Fourier spectra $\zeta_{mn}(t_i)$: 
hexagons at $t_1$,
beaded stripes at $t_2$, quasi-hexagons at $t_3$ and $t_5$,
nonsymmetric beaded stripes at $t_4$ and $t_6$.}
\label{fig:echantillons}
\end{figure*}

Starting from zero velocity and an initial random perturbation of the flat
interface, our simulations produced a hexagonal pattern which oscillates
subharmonically with the same spatio-temporal spectrum as
\cite{KEMW2005,*KEMW2009}, as detailed in \cite{Perinet_jfm_2009}.  
In the experiments, hexagons are
transient and difficult to stabilize \cite{Wprivate}, competing with squares and 
disordered states \cite{KEMW2005,*KEMW2009,WMK2000,*WMK2003}.
In our simulations, 
after about 10 subharmonic periods, we observed a drastic
departure from hexagonal symmetry.
Figure \ref{fig:deltazeta_complet} shows the 
instantaneous maximum height $\max_{x,y}\zeta(x,y,t)$ and its envelope
$\max_{x,y,[t,t+T]}\zeta(x,y,t)$.
Surrounding the time-evolution plot are contours
of the interface height at representative times
over one subharmonic cycle.

At times $t_1+jT/4$, the patterns are hexagonal.  Each is invariant under
rotations by $\pi/3$ and under reflection, for example about $y=n\lambda_c$;
these operations generate their isotropy subgroup (group of symmetries), which
is isomorphic to $D_6$. We call the patterns at $t_2+jT/4$ beaded stripes.
They satisfy the two symmetry relations:
\begin{equation}
\zeta(x,n\lambda_c-y)=\zeta(x,y) =
\zeta(m\lambda_c/\sqrt{3}+\tilde{x}_0-x,y+n\lambda_c)
\label{eq:pmg}\end{equation}
where $\tilde{x}_0\approx\lambda_c/(2\sqrt{3})$ is a spatial phase.
The second equality in \eqref{eq:pmg} is variously called
shift-and-reflect or glide-reflection symmetry. These invariances describe the
crystallographic group called pmg or p2mg \cite{WMK2000,*WMK2003,wallpaper}, which is
isomorphic to $Z_2\times Z_2$. At later times, the patterns have no exact
symmetries.  Nevertheless, the patterns at times $t_3+T/4$ and $t_5+3T/4$
contain large flat cells surrounded by six small peaks
like those at $t_1+3T/4$, which lead us to call them
quasi-hexagons.  Quasi-hexagons at $t_3$ and $t_5$ appear in two distinct
forms, which are related by the spatio-temporal symmetry
\begin{equation}
\zeta(m\lambda_c/\sqrt{3}+x_0-x,y+y_0,t_3+T/2)=\zeta(x,y,t_5)
\label{eq:stsym}\end{equation}
with phases $x_0\approx0.7\lambda_c/\sqrt{3}$ and
$y_0\approx \lambda_c/2$.  The patterns at $t_4$ and $t_6$ resemble those at
$t_2$ but do not satisfy \eqref{eq:pmg}; we call these nonsymmetric beaded
stripes.  These are also related by the spatio-temporal symmetry
\eqref{eq:stsym}.

\begin{figure*}
\begin{minipage}[h!]{5.9cm}
a)\includegraphics[width=6.1cm]{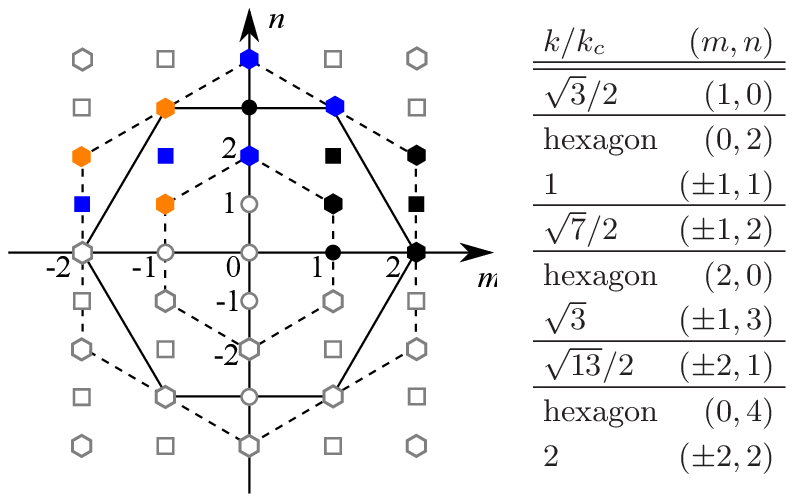}
\end{minipage}
\begin{minipage}[h!]{9.8cm}
c)\includegraphics[width=8.2cm,clip]{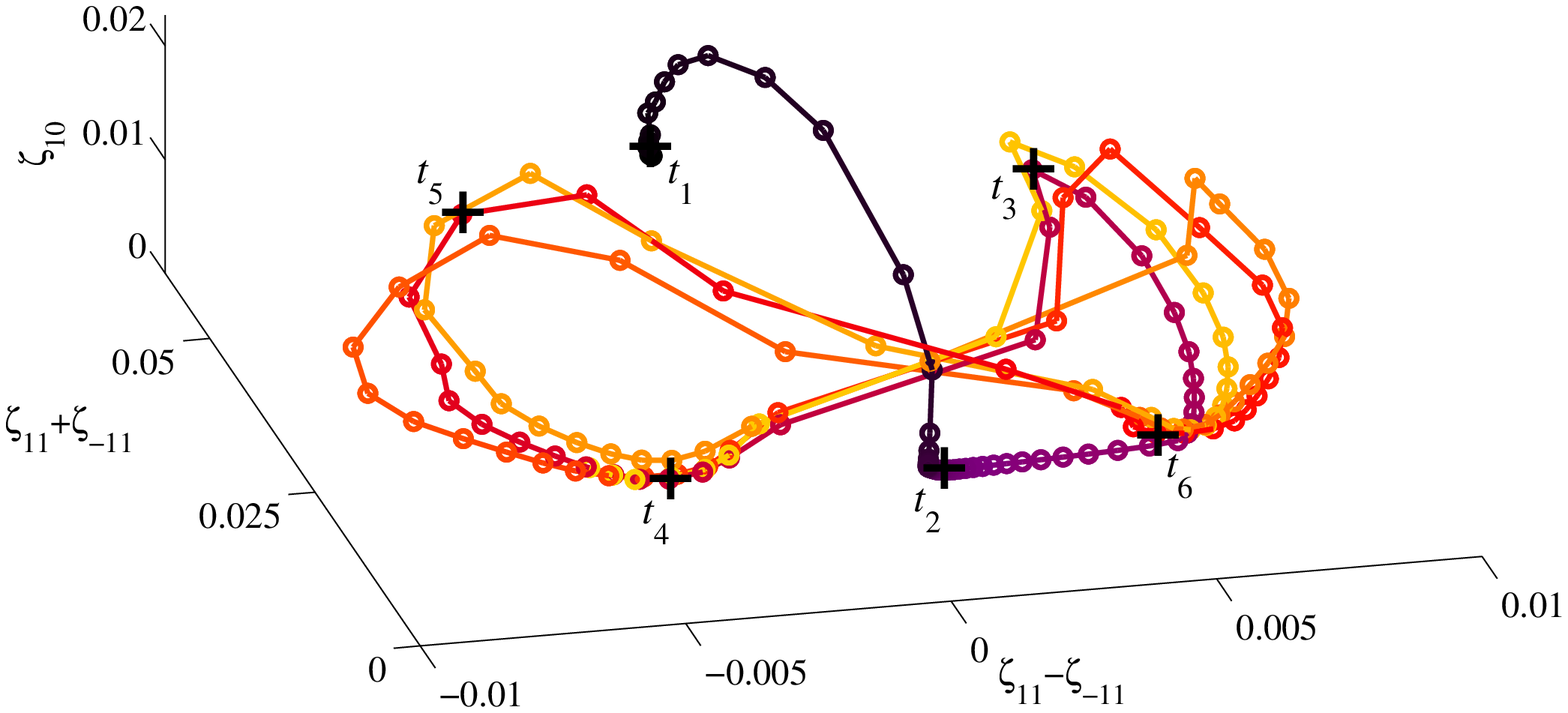}
\end{minipage}
\begin{minipage}[h!]{18cm}
\includegraphics[width=17.4cm]{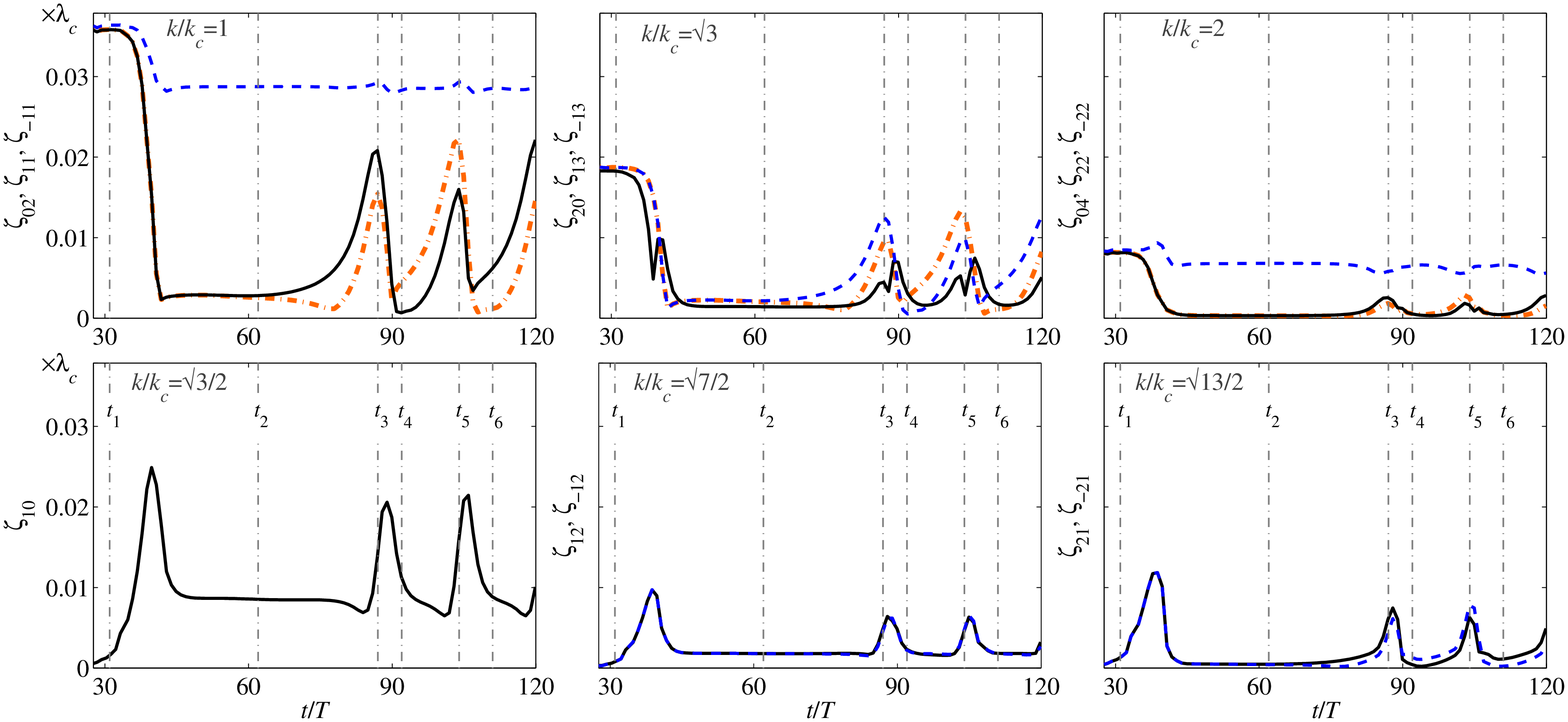}
\vspace*{-1cm}\flushleft b)
\end{minipage}
\caption{a) Spatial Fourier grid of the domain. Hexagonal symbols: sets of six
  $\bk_{mn}$ which share the same length $k$.  This portion of the grid
  contains three hexagons, with $k/k_c=1$, $\sqrt{3}$ and $2$.  The hexagonal
  modes are located at the vertices (and not the sides) of the large hexagons
  indicated by solid and dashed lines.  Square symbols: sets of four
  $\bk_{mn}$ which share the same $k$.  Circles: other modes. Colors
  differentiate between $\bk_{mn}$ and its images under rotations of $l\pi/3$
  for hexagonal modes or under reflections across the $n$ axis for the others.
  b) Temporal evolution of $\zeta_{mn}(t)$, grouped by length $k$.  The color
  code is that used in the spatial Fourier grid. Dashed blue curves: hexagonal
  modes $(0,2)$, $(1,3)$, $(0,4)$ and non-hexagonal modes $(-1,2)$,
  $(-2,1)$. Solid black curves: hexagonal modes $(1,1)$, $(2,0)$, $(2,2)$ and
  non-hexagonal modes $(1,0)$, $(1,2)$, $(2,1)$. Dash-dotted orange curves:
  hexagonal modes $(-1,1)$, $(-1,3)$, $(-2,2)$. c) 3D phase portrait of
  temporal evolution of modes for an extended simulation.  Points are equally
  spaced in time; their color evolves in time from dark to light. Times
  $t_1\ldots t_6$ marked by crosses and by vertical lines in b).}
\label{fig:spectres_complet}
\end{figure*}

To quantify the behavior in Fig.~\ref{fig:deltazeta_complet},
we have studied the spatial Fourier spectrum.
The rectangular box
of Fig.~\ref{fig:domaine_complet} constrains the wavevectors to lie on
the grid $\bk_{mn} \equiv\left(\sqrt{3} m \be_x+n\be_y\right)k_c/2$,
$m,n\in\mathbb{Z}$, as shown in Fig.~\ref{fig:echantillons} and
\ref{fig:spectres_complet}a.
We define the time-filtered spatial Fourier transform:
\begin{equation*}
\frac{\zeta(\bx,t)}{\lambda_c}=\sum_{m,n} e^{i \bk_{mn}\cdot \bx}
\hat\zeta_{mn}(t)\; ,\qquad
\zeta_{mn}(t)\equiv\max_{[t,t+T]}|\hat{\zeta}_{mn}(t)|
\end{equation*}
In Fig.~\ref{fig:echantillons}, we plot 
$\zeta_{mn}(t_i)$ for $t_1\ldots t_6$.
For the hexagonal pattern at time $t_1$ the modes with
non-negligeable amplitude are those belonging to hexagons,
primarily $(m,n)=(\pm 1,\pm 1)$ and $(0,\pm 2)$, with $k/k_c=1$, 
but also $(\pm 2,0)$ and $(\pm 1,\pm 3)$, with $k/k_c=\sqrt{3}$, 
and $(0,\pm 4)$ and $(\pm 2, \pm 2)$, with $k/k_c=2$.
The spectrum at $t_2$ is dominated by modes 
$(0,\pm 2)$ and $(\pm 1,0)$, which combine to form the 
beaded striped patterns -- with 
one bead over $L_x$ and two stripes over $L_y$ -- 
seen in Fig.~\ref{fig:deltazeta_complet} at $t_2$.
The spectra at $t_3$ and $t_5$ combine wavenumbers seen at $t_1$ and $t_2$, 
while those at $t_4$ and $t_6$ are asymmetric versions of that at $t_2$.
The spatio-temporal symmetry \eqref{eq:stsym}
is manifested by $\zeta_{-m,n}(t_{3,4})=\zeta_{mn}(t_{5,6})$.

Figure \ref{fig:spectres_complet}b shows the time evolution of 
$\zeta_{mn}(t)$ for the dominant wavevectors,
grouping those with the same $k$ value into a single graph.
Because $\zeta_{m,-n}=\zeta_{-m,n}$ (resulting from the reality condition
$\hat{\zeta}_{m,-n}=\hat{\zeta}^\ast_{-m,n}$), we plot only positive $n$.
We first describe the evolution of the four most dominant modes in the 
leftmost column. 
During the hexagonal phase at $t_1$, the only modes with non-negligeable
amplitude are those belonging to hexagons.  At $t/T\approx 40$, these drop
abruptly, with the exception of $\zeta_{02}$. This is accompanied by a burst in
amplitude of some of the non-hexagonal modes, notably $\zeta_{10}$, followed
by its saturation.
Shortly after $t_2$, $\zeta_{1,1}$ begins to rise, 
followed by $\zeta_{-1,1}$; eventually $\zeta_{10}$ rises as well;
we conjecture that its growth is fueled by 
mode interactions arising from $\bk_{11}+\bk_{-1,-1}+\bk_{1,0}=\bk_{10}$.
By $t_3$, these have attained values which 
approach $\zeta_{02}$, leading to the quasi-hexagonal patterns at $t_3$.
Mode amplitudes $\zeta_{\pm 1,1}$ then fall quickly, followed by $\zeta_{1,0}$,
leading again to a short-lived beaded striped pattern at time $t_4$.  The
cycle then repeats, but this time $\zeta_{-1,1}$ rises before $\zeta_{1,1}$
and attains a higher peak at time $t_5$, leading to the 
difference between the quasi-hexagonal patterns at $t_5$ and $t_3$. The next
cycle shows $\zeta_{11}$ leading again at $t_6$.
Most of the higher $k$ modes behave like their lower $k$ analogs.

Figure \ref{fig:spectres_complet}c shows a phase portrait, projecting the
dynamics onto coordinates, $\zeta_{11}+ \zeta_{-1,1}$,
$\zeta_{11}-\zeta_{-1,1}$ and $\zeta_{10}$ to represent the dynamics.  Here,
the simulation has been extended and shows several additional repetitions.
The concentration of points indicate that the hexagonal pattern at time $t_1$
and the beaded striped pattern at $t_2$ are saddles.  Afterwards, the
trajectory consists of two crossed loops connecting $t_3$, $t_4$, $t_5$ and
$t_6$.  Several dynamical-systems scenarios lead to limit cycles which visit
symmetrically related sets, e.g.  Hopf bifurcations \cite{CK1994} or
heteroclinic cycles \cite{BGP2000}.
Since the saddles at $t_1$ and $t_2$ are not part of 
the periodic cycle, nor its center, our planned investigation of its origin 
will be challenging. 

A complete analysis of the general bifurcation problem on a hexagonal lattice
\cite{BG1983,GSK1984} shows that the patterns at onset are
hexagons and stripes.
This analysis of steady states on a hexagonal grid does not apply 
directly to our study of subharmonic standing
waves on the rectangular lattice of Fig.~\ref{fig:domaine_complet}. 
At the other extreme, in large-domain experiments, hexagonal patterns are often 
observed to undergo other spatio-temporal dynamics, such as 
competition with squares \cite{KEMW2005,*KEMW2009,WMK2000,*WMK2003,AWR2000}, 
which are inaccessible to current mathematical analysis and to our simulation. 

We have observed complex long-time temporal behavior in a fully resolved
three-dimensional simulation of Faraday waves in the minimal domain 
which can accomodate a hexagonal pattern.  
Although this scenario may prove to be replaced by other dynamics
in large domains, we believe that it is of interest in its own right 
and that it may well be applicable to other pattern-forming systems. 

We thank P.-L. Buono and M. Golubitsky for sharing their knowledge of 
symmetry and C. Wagner for in-depth remarks on experiments. 
N.P. was partly supported by the 
Natural Sciences and Engineering Research Council of Canada.

\vspace*{-1cm}
\bibliography{biblio}

\end{document}